\documentstyle[11pt,newpasp,epsf]{article}
\begin{document}

\title{Mergers and star formation in SPH cosmological simulations}
\author{Patricia B. Tissera}
\affil{Instituto de Astronom\'{\i}a y F\'{\i}sica del Espacio, Argentina}

\begin{abstract}
The star formation rate history of galactic objects in 
hydrodynamical cosmological simulations
are analyzed in relation to their merger histories. The findings suggest that massive mergers produce more efficient
starbursts and that, depending on the internal structure of the objects, double starbursts could also occur.
\end{abstract}

Our knowledge of the Universe has dramatically improved in the last decade thanks to 
the large amount of information gathered  on nearby and high redshift objects that
makes possible to attempt tests of theoretical models on more realistic basis.
Galaxies surveys   have been produced and used to try to 
test structure formation scenarios and cosmological models.
However,  a main problem in these studies is the relationship between galaxies and the underlying overall
mass distribution: do galaxies trace the mass? or how galaxies trace the mass?
In particular, it is now known that clustering properties of galaxies depend, at least, on their color and luminosities.
And that, selection effects and galaxy formation conspire to bring different types of galaxies into the samples
as function of the $z$. 
Hence, understanding the clustering properties at different $z$ implies being able to explain how galaxy formation 
proceeds within dark matter halos. At high $z$, Lyman Break Galaxies (LBG) have provided us with information 
of objects in the early phase of evolution of the Universe. These objects have been found to be strongly clustered.
 The good
agreement found between both abundance and clustering properties of massive dark matter halos and typical
LBGs have led to support the idea that star formation  is mainly determined by gas accretion onto the potential well
of halos  (i.e., Aldelberger et al 1998).
However, assuming a hierarchical clustering scenario for the formation of the structure
brings mergers and interactions into play as common
events in the  formation and evolution of galaxies, making the picture more complex.

The observed astrophysical properties of galactic objects depend  on their star formation
 (SF) histories. Since
  SF is a complex mechanism regulated by different physical processes acting at different scales such as
supernova feedback, gas cooling, interactions, mergers, disk instabilities, etc.,  its link with mass
may be  not straightforward.
Violent event may play an important role in the SF histories of
a galactic object. Indeed, observations of nearby and high-z objects suggest an increment of both, the 
SF activity and the rate of mergers with $z$.
 From a theoretical point of view, 
several authors have studied their effects on the SF history
of galaxies by using prepared merger simulations (i.e., Mihos \& Hernquist 1996; Barnes \& Hernquist 1996) or
introducing their effects  in 
semianalytical models (Kauffman et al. 1993; Somerville \& Primack 1998; Percival \& Miller 1999).
In this work, I present results on a detail analysis of the star formation history of typical galaxy-like objects
in relation to their merger history in hydrodynamical simulations within a cosmological context.

\section{Simulations}
The simulations analyzed in this work are described in details by Tissera (1999). Briefly, they
take into account the gravitational and hydrodynamical evolution of the matter
including an algorithm to transform  cold and dense gas into stars.
All experiments are consistent with a standard Cold Dark Matter universe. The simulated boxes
have $5 h^{-1}$ Mpc length with $N=64^3$ total particles. Baryonic particles represent 10 $\%$
of the total mass. Note that dark matter and baryonic particles have the same mass,
$M_{\rm part}= 2.6 \times 10^8 {\rm M{_\odot}}$.
We run three simulations, S.1, S.2 and S.3. Simulations S.1 and S.2 share the same initial conditions
while S.3 is a different realization. The bias parameter $b$ for S.1 and S.2 is $b=2.5$, and for S.3, $b=1.67$.

The star formation algorithm is based on the Schmidt law: $\dot \rho_{\rm star}=-c\rho_{\rm gas}/t_*$,
 where $c$  is the
star formation efficiency ($c=0.01,0.1, 0.01$ for
S.1, S.2 and S.3, respectively) and $t_{*}$ is a characteristic time-scale
assumed to be equal to the dynamical time of the particle. There is another parameter $T_{\star}$ which
is the critical temperature for gas particles to be transformed into stars:
 $T_{\star}= 10^4$ K and $3 \times 10^4$ K
for S.1 and S.2, and S.3 respectively. The values of $c$ and $T_{\star}$ together with $b$ affect
the SF in the simulated box. Globally, simulations S.1, S.2 and S.3  have transformed $12 \%$, $28 \%$ and
$7.5\%$, respectively, of their total baryonic mass into  stars at $z=0$.
 The SF
parameters used in S.1 and S.3 allow the transformation of gas into stars only in the very dense regions.
In practice, this fact implies that gas particles have to get to the center of the progenitor before can be
converted into stars. Supernova feedback has not been included in these simulations.

\section{Mergers and Star Formation History}

Galactic-like objects (GLO) are identified at their
virial radius at $z=0$ and then, followed
back in time. Their merger trees are reconstructed by
tracking back all particles that belong to a given GLO
at $z=0$, and the progenitor object is chosen as
the more massive clump within this merger tree. We consider mergers with clumps larger than 
$10 \%$ the progenitor viral mass at the time of the merger.
We will define a merger event as the whole process from the time
the two objects are first identified to share the same dark matter halo,
to the time there two baryonic clumps mergerd. During the
orbital decay phase, the minor colliding baryonic clump
will be refereed as  the satellite.
\begin{figure}
\plotfiddle{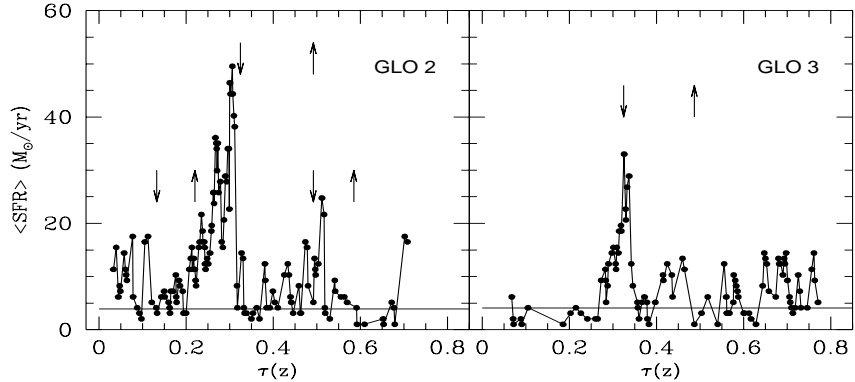}{2in}{0}{60}{50}{-190}{-210}
\caption{Star formation history of galaxy-like objects in simulation S.1 as a function of look-back
time. Merger events are indicated by arrows }
\end{figure}
It has to be stressed that we are only going to study
those GLOs with more than 250 baryonic particles inside the main object
and from $z=1$, to diminish as much as possible numerical resolution
effects. 

The SF rate history of each GLO is estimated by reckoning the stellar mass formed in its progenitor objects at
each $z$ and, then, smoothing these distributions over time. In Fig.1 we show the mean SFR histories
for GLOs 2 and 3 in S.1 as a function of look-back time. The arrows pointing up indicate the times at which
the satellite enters the virial radius of the progenitor, while the actual fusion of the baryonic clumps
is indicated by the arrows pointing down.
As can be seen from this figure, there is an increase of the SF activity during the merger events. This situation
is common to all GLOs analyzed.

In order to identify  more rigorously the SFR peaks, we estimate the overall minimum SFR in a GLO at any redshift. 
We then subtract a factor $f$ of this minimum from the total SFR history, so that peaks are clearly
identified as the values with a signal larger than a threshold $\delta_{\rm min}$ (solid line in Fig.1).
 We took $f=3$ for  all GLOs, which yields mean values of  $<\delta_{\rm min}>=0.90, 1.45, 0.89 {\ \rm M_{\sun}/yr}$
for GLOs in S.1, S.2 and S.3, respectively.  Hence, all SFR history can be described as the contribution of
two components: a quiescent  SF and a series of starbursts (SB).
From all SBs identified, we are only concern with those that occurred within a merger event.
Once the starbursts are isolated, it is direct to define their height, $\sigma_{\rm star}$,
the total stellar mass formed, $M_{\rm burst}$, and their duration, $\tau_{\rm burst}$.
Table 1 shows their mean values in units of 
 ${\rm M_{\odot}/yr}$, $10^{8} {\rm yr}$ and ${\rm 10^{10} M_{\odot}}$, respectively.
These quantities determine the characteristics of each starburst together with the
gas richness of the system, $M_{\rm star}/M_{\rm bar}$, and the virial mass ratio of
the colliding  objects, $M_{\rm sat}/M_{\rm pro}$, at the time of the merger.
$M_{\rm star}$ and $M_{\rm bar}$ are the stellar and baryonic content of the
merging systems; $M_{\rm sat}$ and $M_{\rm pro}$ are the virial masses of the
satellite and progenitor, respectively.

\begin{table}

\footnotesize
\caption{Mean values for the parameters of the stellar bursts. }
\begin{tabular}{crrc}

S & $<\sigma_{\rm star}>$   & $<\tau_{\rm burst}>$ & $< M_{\rm burst}>$ \\
1& 20.79  &9.28  &1.36\\
2& 40.02  &11.49  &1.79\\
3& 12.51  &5.14  &0.39\\

\end{tabular}
\end{table}

An exhaustive analysis of the properties of the SBs yields the following conclusions.
We found no correlation between $\sigma_{\rm star}$ and $M_{\rm sat}/M_{\rm pro}$. 
Mergers with equal $M_{\rm sat}/M_{\rm pro}$  are found to produce SBs of different strengths. 
A study of   $\sigma_{\rm star}$ versus the gas richness of the systems defined by  $M_{\rm star}/M_{\rm bar}$
shows no correlation  indicating that not always the more gas-rich object have the larger $\sigma_{\rm star}$
and that equal gas-rich GLOs have different  $\sigma_{\rm star}$, even within the same simulations. 
Hence, the fact that equal massive mergers produce different    $\sigma_{\rm star}$ bursts in the same
simulations can not be directly linked to a difference in the gas richness of the systems.       

We also found that   $\sigma_{\rm star}$ shows no correlation signal  with neither $M_{\rm gas}$ nor $M_{\rm pro}$.
Distinguishing between minor and major mergers ($M_{\rm sat}/M_{\rm pro}=0.35$) suggests that minor mergers
can trigger bursts as strong as   major ones even if the systems are less gas-rich.
On the other hand, no correlation signal was found between neither $M_{\rm burst}$
 and $M_{\rm gas}$ nor  $M_{\rm burst}$   versus  $M_{\rm star}/M_{\rm bar}$.
However, a trend was found for major mergers to be more efficient at transforming gas into stars during
mergers. This fact  can be seen in Fig.2a where we plot SB efficiencies $M_{\rm burst}/M_{\rm gas}$ versus  
 $M_{\rm sat}/M_{\rm pro}$ for bursts in S.1 (circles), S.2 (triangles) and S.3 (pentagons). 
Mean  values of SB efficiencies  for minor mergers ($M_{\rm sat}/M_{\rm pro} <0.35$)
are 0.24, 0.23 and 0.10 in S.1, S.2 and S.3 respectively, while for  major ones:  0.35, 0.72
and 0.19, respectively.

During the inspection of the starbursts, it was also found that during certain mergers events, two double peaks were
identified: a first one during the orbital decay phase of the satellite, and, a second one during the 
fusion of the baryonic cores. 
As have been reported by  Mihos \& Hernquist (1996) and  Dominguez-Tenreiro, Tissera, S\'aiz (1998) among others,
the presence of the first SB could be related to the stability properties of the gaseous disks. This authors
found that the presence of a compact stellar bulge assures the stability of gaseous disks during violent events
preventing the triggering of strong gas inflows during the orbital decay phase. These tidal induced inflows
could   be related to the presence of the first SB (see
Tissera et al. 2000 for details). In this sense,
 we look for a correlation between the presence of these
double starbursts and the formation of a compact stellar mass concentration (i.e., as an indicator of 
the formation of a compact stellar bulge). We measured the heights of the bursts and ratio 
between the stellar content of the progenitor at the time the satellite enters its virial radius and the progenitor total
stellar mass at $z=0$ ($M_{\rm star}^{z}/M_{\rm star}^0$).
 This ratio
gives a rough idea of the presence of a well-formed stellar mass concentration at the center.
Fig.2b shows this relation. As can be clearly seen, we only detect double bursts in those GLOs
with  $M_{\rm star}^{z}/M_{\rm star}^0 < 0.40$. All GLOs with $M_{\rm star}^{z}/M_{\rm star}^0 >0.40$ have
only single bursts when the fusion of the baryonic clumps occurs.
 It has to be mentioned that $\sigma^{1}_{\rm star}/\sigma^{2}_{\rm star}$ does not correlate with either,
$M_{\rm star}/M_{\rm bar}$ or $M_{\rm gas}$. 
Looking at the starburst efficiencies of these double events, we found a trend for first SB to
be more efficient than single ones. The mean value  for mergers in S.1, S.2 and S.3 are: (0.39,0.08), (0.56,0.39) and
(0.16,0.03) for stabursts with $M_{\rm star}^{z}/M_{\rm star}^{0}$ smaller and greater than 0.40, respectively
(Tissera et al. 2000).

\section{Summary}

Using hydrodynamical simulations in a cosmological context has allowed us to consistently study
the star formation and merger histories of galactic objects. Our findings suggest that,
if a hierarchical clustering scenario is adopted, mergers and interactions play a critical role in the
regulation of star formation since they are able to violently compressed the gas in short time-scales.
Given the Schmidt law, any increase in the gas density would lead to an increase in the SF rate. Hence, how baryons
distribute within the dark matter halos  seems to be relevant for the SF process. Even more,  how the systems
behave during violent events could also depend on its internal structure  might have a non-negligable effect on SF. And, since properties of galactic
objects change with  time as they evolve, how the same physical mechanisms affect  star formation
might  also change with $z$ making it difficult to stablish a direct link between mass and star formation.

\begin{figure}
\plotfiddle{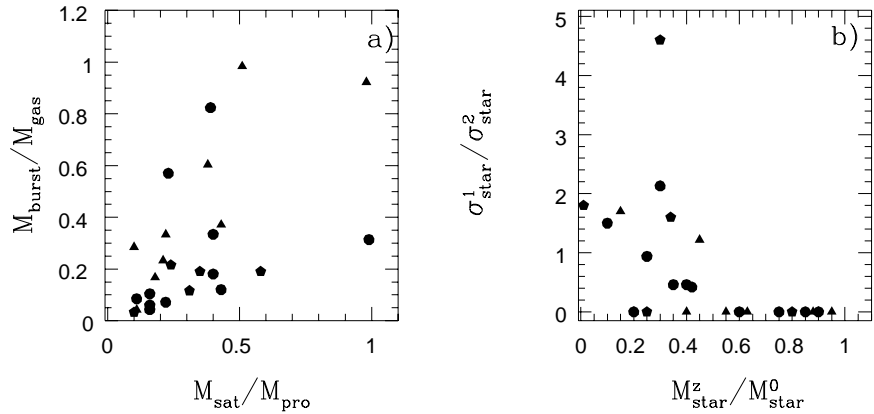}{2in}{0}{60}{60}{-190}{-265}
\caption{a)Starburst efficiency versus the relative virial masses of the colliding systems. A trend is found
for more massive mergers to produce more efficient bursts. 
b) Ratio between the strengths of the
double star bursts ($\sigma_{\rm star}^{1}/\sigma_{\rm star}^{2}$)
and the fraction of stars already formed  in the
progenitor  ($M_{\rm star}^{z}/M_{\rm star}^{0}$)
at the $z$ of the merger ($z_1 > z_2 $).
Single bursts have been assigned
$\sigma_{\rm star}^{1}/\sigma_{\rm star}^{2} =0$.}
\end{figure}

\acknowledgements The author thanks the R. Dominguez-Tenreiro, A. S\'aiz and C. Scannopieco for
allowing the inclusion of some common results to be published in Tissera et al. (2000).

\end{document}